\documentclass[a4paper]{article}
\usepackage{graphicx}

\begin{document}

\begin{center}
{\bf \Large
Generic behaviours in  impact fragmentation}
\bigskip

{\large
N. Sator\footnote{sator@lptmc.jussieu.fr}$^{\star}$,
S. Mechkov$^{\star \ddag}$
and
F. Sausset$^{\star}$
}
\bigskip

\bigskip
$^\star${ Laboratoire de Physique Th\'eorique de la
  Mati\`ere Condens\'ee, Universit\'e Pierre et Marie Curie (Paris 6),
  UMR CNRS 7600
  - 4 place Jussieu, 75252 Paris Cedex 05, France}, 

$^\ddag${ Laboratoire de Physique Statistique, Ecole Normale
  Sup\'erieure - 24 rue Lhomond, 75231 Paris Cedex 05, France}

\bigskip
\today
\end{center}

\begin{abstract}
  We present a simple numerical model for investigating the general
  properties of fragmentation. By use of molecular dynamics
  simulations, we study the impact fragmentation of a solid disk of
  interacting particles with a wall. Regardless of the particular form
  of the interaction potential, the fragment size distribution
  exhibits a power law behaviour with an exponent that increases
  logarithmically with the energy deposited in the system, in
  agreement with experiments. We expect this behaviour to be generic
  in fragmentation phenomena.
\end{abstract}

\noindent
{\em PACS numbers:} 46.50.+a, 62.20.Mk, 89.75.Da

\noindent
{\em Keywords:}  fragmentation, cracks, scaling laws

\bigskip

\section{Introduction}

From atomic nuclei to supernovae, including plates and rocks, every
cohesive system can be broken into fragments, provided that the
deposited energy is sufficiently large compared to its cohesive
energy~\cite{beysens95}. The ubiquity and the fundamental role of
fragmentation are reflected in planetary formation~\cite{matsui77} and
geological phenomena~\cite{turcotte86} like explosive volcanic
eruptions~\cite{kaminski98}, as well as in the determination of the
phase diagram of nuclear matter~\cite{elliott03,campi00}. Furthermore,
a major part of activities in building and pharmaceutical industries
makes use of particle size reduction
processes~\cite{beysens95,frances98}.

In spite of this wide variety of length scales and materials,
fragmentation processes present some generic features. The most
striking of them is the frequently observed power law form of the
fragment size distribution, whose exponent $\tau$ varies from 1 to
about 2.5 according to the data provided by
experiments~\cite{beysens95,turcotte86,elliott03,campi00,meibom96,ching99,oddershede93,kadono02,moukarzel07,matsui82,ishii92,kadono97,arakawa99,piekutowski95,katsuragi05,wittel04}
and natural phenomena~\cite{turcotte86,kaminski98,oddershede98}.

Over the last decades, much effort has been devoted to understanding
the origin of this scaling behaviour and a number of proposals have
been advanced. The scattering of the values of $\tau$ casts some
doubts~\cite{meibom96,ching99} on a self-organized critical mechanism
proposed by Oddershede {\it et al.}~\cite{oddershede93}, but raises
the question of the existence of universality classes in
fragmentation~\cite{kun99,astrom00}. Thus, collision experiments of
atomic nuclei exhibit a percolation behaviour of the fragment size
distribution ($\tau \simeq 2.2$)~\cite{campi00}, which is typical of
the fragmentation of supercritical fluids~\cite{sator03}. Likewise,
evidence for criticality has been found in hydrogen cluster
fragmentation~\cite{farizon98}. In contrast, the definition of
universality classes in fragmentation of macroscopic systems is still
an open question.

As matter stands, fragmentation may depend on several parameters like
the dimension of the space, the shape, size and material of the
fragmenting object, as well as on the way it is broken and on the
amount of energy deposited in it.  Simple analytical models based on
rate equations~\cite{boyer95}, binary fissions~\cite{meibom96}, or
sequential fragmentation~\cite{kadono02,astrom04a} although elegant
and instructive have limitations and cannot take into account all the
above-mentioned parameters. As an example, most of the analytical
models predict a single exponent $\tau$ depending only on the
dimension of the space, which experiments seem to refute.

Molecular dynamics calculations have been proven to be a powerful
technique for addressing the role played by the parameters that may
influence fragmentation processes. Furthermore, simulations allow one
to study the dynamics of an irreversible and strongly
out-of-equilibrium process like fragmentation. Until now, mainly two
kinds of numerical experiments have been proposed. On the one hand,
explosive fragmentation of a system is simulated by initially
allocating a centrifugal velocity to the particles, which interact
through a Lennard-Jones
potential~\cite{ching99,astrom00,diehl00,araripe05}.  On the other
hand, various numerical simulations are concerned with the impact
fragmentation of brittle
solids~\cite{kun99,astrom00,myagkov05,behera05,thornton96}. Of
particular interest, Kun and Herrmann proposed a sophisticated model
that takes into account elastic, shear and torque interactions between
particles of a two-dimensional disk which impacts another
disk~\cite{kun99} or a hard plate~\cite{behera05}. They suggested that
fragmentation of solids occurs as a continuous phase transition from a
damaged state to a fragmented state when the energy deposited in the
system increases.

In the same spirit, we propose a simple model providing a general
framework for investigating the role of the different parameters that
may be relevant in fragmentation processes. In this paper, we study
the impact fragmentation of a two-dimensional disc of interacting
particles with a wall, as a function of the energy deposited in the
system. The role played by the materials (interaction potential) and
by the size of the disc (number of particles) are especially
investigated.

Note that although fragmenting systems are usually three-dimensional,
some experiments are performed on platelike objects with a
two-dimensional experimental
setup~\cite{kadono02,kadono97}. Furthermore, we are here concerned
with general problems that may be tackled by two-dimensional
simulations.

\section{Model}

To provide a generic frame of reference, the fragmenting system is
made up of particles which represent mesoscopic grains of materials,
without specifying their exact nature.  These particles interact
through a two-body central Lennard-Jones type potential:
\begin{equation} 
\label{eq.1} 
\mathrm{v(r_{ij})}=v_0 \epsilon
\bigg[\bigg(\frac{\sigma}{\mathrm{r_{ij}}}\bigg)^{a}-\bigg(\frac{\sigma}{\mathrm{r_{ij}}}\bigg)^{b}\bigg],
\end{equation}
where $\mathrm{r_{ij}}$ is the distance between particles $i$ and $j$, and the
two constants, $\epsilon$ and $\sigma$, set the energy and length
scales respectively.  Hence, $\sigma$ is typically the diameter of the
particles. The two exponent parameters, $a$ and $b$, control the range
of attraction and $v_0$ is used to set the minimum of the potential to
$-\epsilon$. In order to mimic the cohesive interaction at a
mesoscopic scale, we choose a very short-range potential
($v_0=107.37$, $a=80$ and $b=78$) with a typical range of attraction
as short as $0.1\,\sigma$, in addition to the particle diameter. As we
will see in the following, our results are not sensitive to the
particular choice of the parameters $v_0$, $a$ and $b$, that is to the
range of attraction.

At first, a disc of a given number of particles is cut in a
two-dimensional triangular lattice with a lattice spacing
corresponding to the minimum of the interaction potential given by
eq.~(\ref{eq.1}). The fragmenting disc is thus a perfect crystal
(without any quenched disorder) prepared at zero temperature and at
the corresponding number density of $\rho \sigma^2 \simeq 1.16$.
Next, the disc is rotated by a random angle and launched towards a
wall, as shown in fig.~\ref{fig.1}, by assigning to each particle a
given impact velocity $\bf{V}$, perpendicular to the wall. We use the
magnitude of the impact velocity as a natural control parameter. The
particles of the disc interact with the wall through the repulsive
part of the potential given by eq.~(\ref{eq.1}), that is $v_0 \epsilon
(\sigma/z_i)^{a}$, where $z_i$ is the distance between the wall and
the particle $i$. We checked that our results do not depend on the
precise form of this hard-core potential.

To study the fragmentation process and its outcomes, we perform
molecular dynamics simulations at constant energy using the Verlet
algorithm~\cite{frenkel01}. The time step is $\delta t=0.0005 \, t_0$,
where $t_0=\sqrt{\epsilon \sigma^2/m}$ is the unit of time and $m$ the
particle mass, which ensures the conservation of the total energy
within $0.01\%$ of its average value. In the following, times and
velocities are expressed in units of $t_0$ and $\sigma/t_0$
respectively. As we will see below, at $t=500$ ($10^6$ iterations),
the fragmentation process has already reached a steady state.  The
main results presented in this paper are obtained with a disc of
$N=1457$ particles, but system sizes up to $N=36\: 289$ were used with
a view to determine the finite size scaling properties of the system.
For $N=1457$, the potential energy per particle is $-2.89 \epsilon$,
slightly in excess of the bulk value ($-3 \epsilon$), because of
surface effects.

During the fragmentation process, fragments are identified as being
self-bound clusters of particles~\cite{sator03,hill55}. More
precisely, two particles are linked if their relative kinetic energy
is lower than the absolute value of their interaction energy.  Note
that a definition of clusters based on an arbitrary cutoff distance
provides the same results when the fragments are far away from each
other.

For a given value of the impact velocity, we calculate the number of
fragments made up of $s$ particles. The fragment size distribution,
$n(s)$, is averaged over 1000 runs by uniformly sampling the initial
random angle of rotation. In fragmentation studies, it is usual to
estimate the cumulative fragment size distribution divided by the
fragment size $s$~\cite{oddershede93}:
\begin{equation} \label{eq.2}
  N(s)=\frac{1}{s}\int_{s}^{\infty} n(s') \;ds'.
\end{equation}
If $n(s)\sim s^{-\tau}$, then $N(s)$ exhibits the same, but less
noisy, power law behaviour.

\section{Results}

First we describe qualitatively the collision of the disc with the
wall, as the impact velocity increases. At very low velocity, a
slightly elastic deformation is observed at the contact with the wall,
followed by the rebound of the disc without any damage.  As the
velocity is increased, deformation becomes irreversible and
fragmentation occurs for $V>0.5$.

\begin{figure}
\begin{center}
\includegraphics[scale=0.25,angle=0]{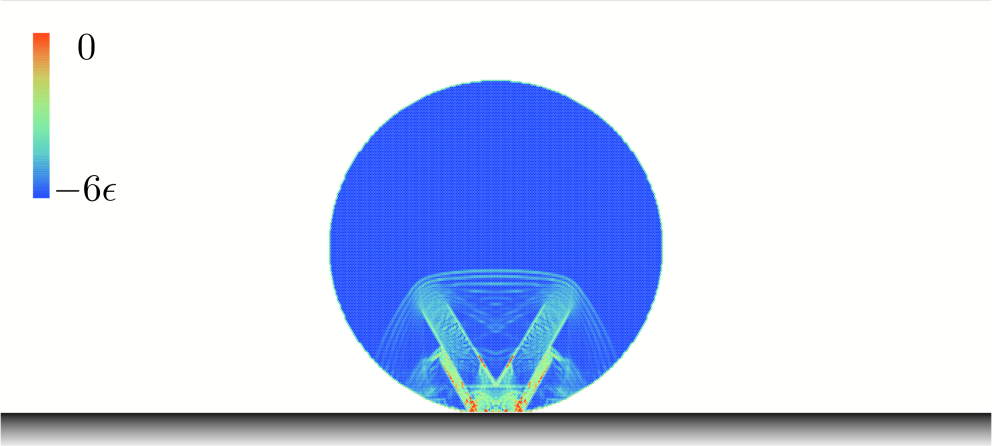}
\vspace{.5cm}
\includegraphics[scale=0.25,angle=0]{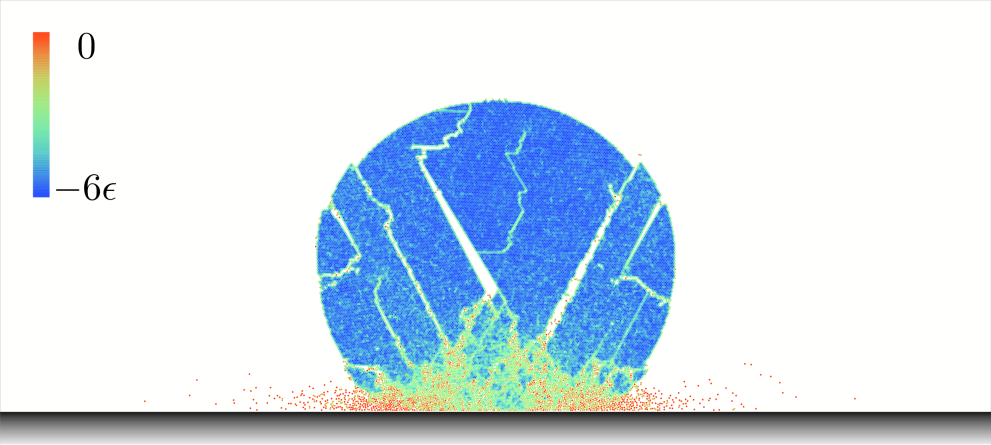}
\vspace{.5cm}
\includegraphics[scale=0.25,angle=0]{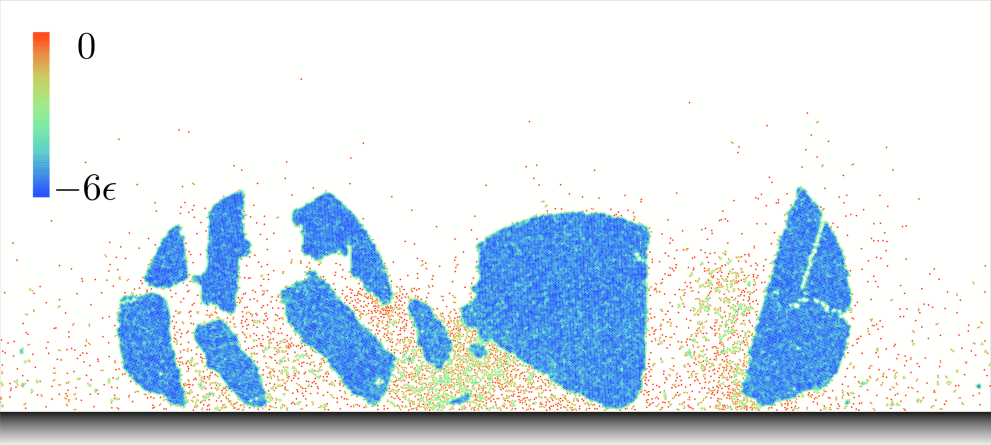}
\end{center}
\caption{(colour online) Snapshots of the fragmentation of a disc of
  $N=36\: 289$ particles launched with a zero initial angle and an
  impact velocity $V=2$, at $t=1$ (top), $t=10$ (middle) and $t=100$
  (bottom).  Particles are coloured according to their potential energy
  from $-6\epsilon$ (blue) to 0 (red).}
 \label{fig.1}
\end{figure}

Three snapshots of a typical fragmentation event are shown in
fig.~\ref{fig.1}. A zero initial angle is chosen in order to clearly
visualize the cracks and the compression waves materialized by the
potential energy of the particles. As can be seen in the top snapshot,
when the disc strikes the wall, a compression wave starts to propagate
from the impact zone through the disc. If the impact velocity is large
enough, the compression wave produces cracks (see the middle
snapshot).  When the cracks coalesce and reach the surface of the
disc, fragmentation takes place. The part of the initial kinetic
energy which is not used to break up the system is mainly transformed
into the internal and kinetic energies of the fragments. As a result
of the collision, fragments fly away from each other, as illustrated
by the bottom snapshot. Note that in spite of the zero initial angle
and the symmetry of the lattice, irregular cracks give birth to rather
rough fragments.  To summarize, the crack pattern we observe is very
much like those found in more sophisticated
simulations~\cite{behera05,thornton96} and in two-dimensional
experiments~\cite{kadono97,kadono02}.

It is interesting to note that the speed of sound,
$c_{\mathrm{sound}}$, in this material is much larger than the impact
velocities investigated in this work ($0.5 <V<4.5$). Indeed, by
observing the propagation of the compression waves (see the top
snapshot in fig.~\ref{fig.1}), we roughly estimate $c_{\mathrm{sound}}
\simeq 100 $.  Furthermore, a calculation of the speed of sound in the
harmonic approximation for a one-dimensional chain of particles
interacting through the potential $\mathrm{v(r)}$ gives the same order
of magnitude: $c_{\mathrm{sound}} = 78$.

In order to better understand the nature of this fragmentation
process, we calculate the mean size of the first and second largest
fragments, denoted $S_{\mathrm{max1}}$ and $S_{\mathrm{max2}}$
respectively. As with percolation theory, we also compute the average
of the second moment of the fragment size distribution, which is
related to the mean fragment size~\cite{stauffer94}:
\begin{equation} \label{eq.3} 
m_2 = \sum_{s} s^{2} n(s),
\end{equation}
where the sum excludes the largest fragment.

\begin{figure}
\begin{center}
\includegraphics[scale=0.35]{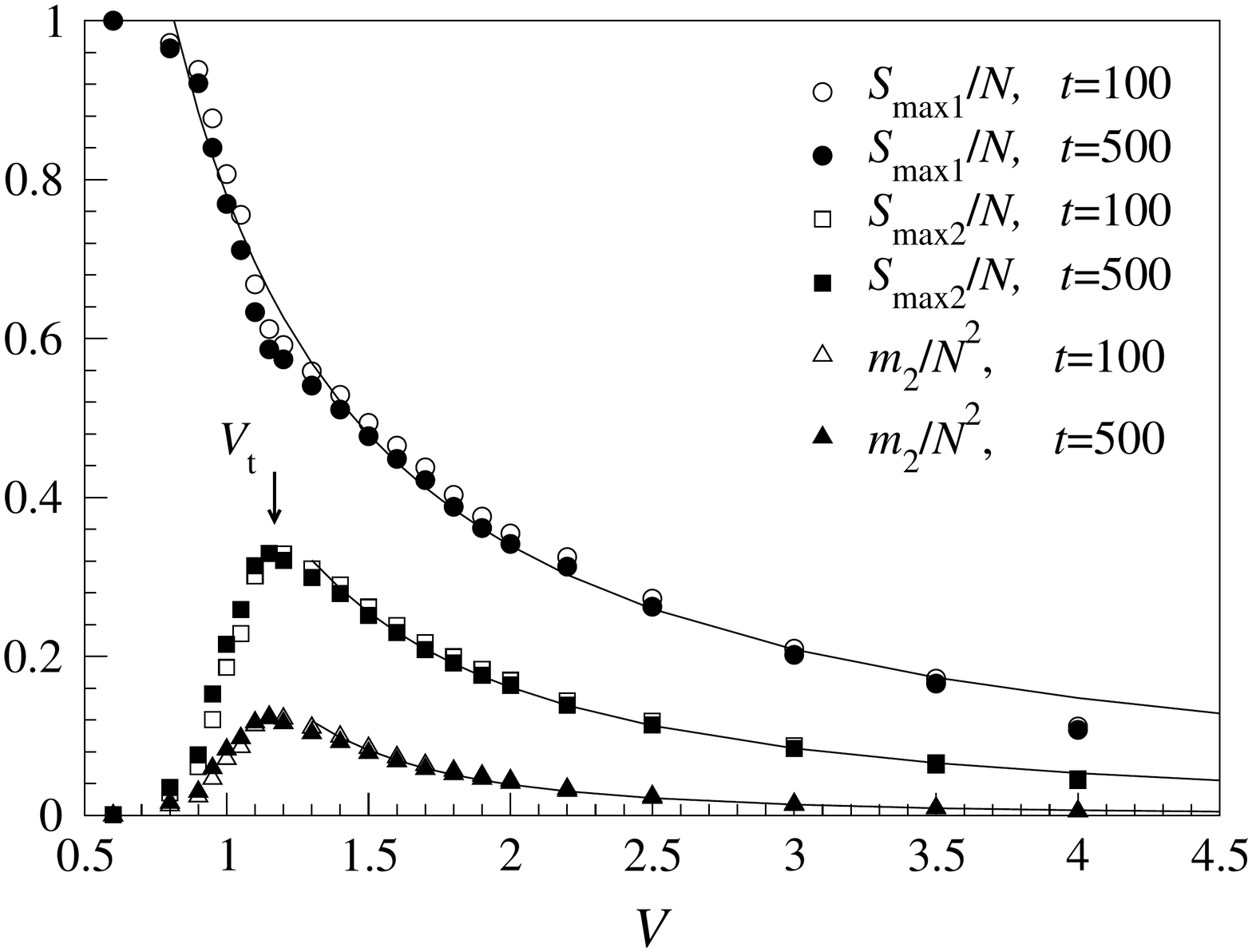}
\caption{The first and second largest fragments and the second moment
  as a function of $V$ for a disc of $N=1457$ particles at time
  $t=100$ (open symbols) and at $t=500$ (solid symbols). The full
  lines correspond to power law fits (see text). The threshold
  velocity is indicated by an arrow.}
\label{fig.2}
\end{center}
\end{figure}

These three quantities are plotted in fig.~\ref{fig.2} as a function
of the impact velocity, at $t=100 $ and at the end of the simulation
at $t=500$. We clearly see that $S_{\mathrm{max1}}$,
$S_{\mathrm{max2}}$ and $m_2$ evolve little between these two
times, showing that the fragmentation process has reached a steady
state. Indeed, the fragment size distribution is rapidly stabilized
after the impact. Although the fragments are heated during the
process, their internal energies are low and allow the evaporation of
only a few particles.

As expected, $S_{\mathrm{max1}}$ decreases with $V$, reflecting the
increasing violence of the impact. On the other hand, both
$S_{\mathrm{max2}}$ and $m_2$ have a maximum at the same threshold
velocity, $V_t =1.15$, as in the simulations performed by Kun,
Herrmann and their collaborators~\cite{kun99,behera05}.  These authors
interpreted this behaviour as a sign of criticality in
fragmentation. We note that the value of $V_t$ decreases slightly with
$N$ (for instance, $V_t\simeq 1$ for $N=36\:289$). As is shown in
fig.~\ref{fig.2}, the decrease of $S_{\mathrm{max1}}$,
$S_{\mathrm{max2}}$ and $m_2$ is well fitted by a power law $\sim
V^{-\gamma}$, with an exponent $\gamma$ equal to 1.2, 1.6 and 2.6
respectively. As can be remarked, in contrast with a critical
behaviour, the threshold velocity does not appear in this power law
form. Moreover, the fit is very good not only in the vicinity of
$V_t$, but also at velocities up to 3.5. It is worth noting that the
power law behaviour of $S_{\mathrm{max1}}$ has been reported in
experimental works dealing with cubic ice
($\gamma=1.8$)~\cite{arakawa99} and aluminum spheres
($\gamma=6.6$)~\cite{piekutowski95}.  Moreover, by fragmenting various
rock types, Matsui and his coworkers~\cite{matsui82} showed that the
exponent seems to depend mainly on the shape of the system: with
$\gamma \simeq 1.4$ for spherical samples and $\gamma \simeq 3$ for
cubic ones.

\begin{figure}
\begin{center}
\includegraphics[scale=0.35,angle=0]{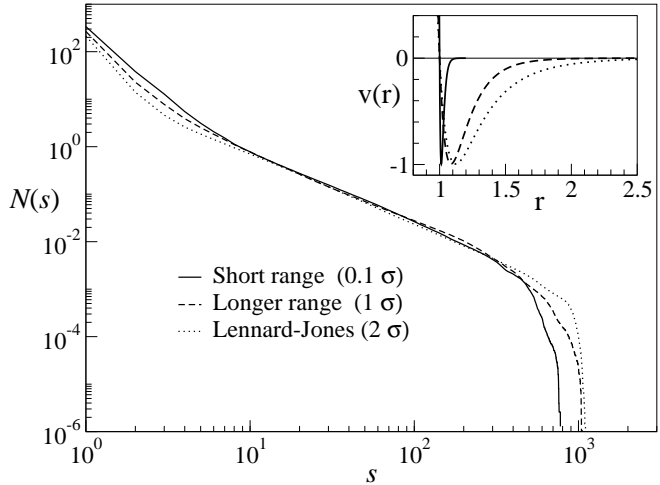}
\caption{Fragment size distributions at $V=2$ for a disc of
  $N=1457$ particles and for three different interaction potentials
  $\mathrm{v(r)}$ plotted in the inset.  The attraction range is given in the
  legend.}
\label{fig.3}
\end{center}
\end{figure}

For $V > V_t$, the fragment size distribution $N(s)$ follows a power
law with an exponent $\tau$ (see fig.~\ref{fig.3}). Of course, the
properties and the number of small fragments ($s \le 10$) may depend
strongly on the particular features of the
system~\cite{kun99,behera05}. However, we are mainly concerned with
the large size range for which generic behaviours are expected.

To test the dependence of our results on the potential interaction, we
carry out simulations with two other potentials having the same form
given by eq.~(\ref{eq.1}), but with a longer attraction range of
$\sigma$ ($v_0=16.29$, $a=13$ and $b=11$) and $2\sigma$ ($v_0=4$,
$a=12$ and $b=6$), the latter being a standard Lennard-Jones potential
(see the inset of fig.~\ref{fig.3}). The fragment size distributions
corresponding to these three potentials are plotted in
fig.~\ref{fig.3}. Strikingly, they share the same slope, and so the
same value of $\tau$, suggesting the robustness of our results.
Differences occur in the small size region and for the largest
fragment, the size of which increases with the range of the potential,
as well as the cohesion of the disc. Furthermore, this result implies
that fragmentation may be, to a certain extent, independent of the
particular material as suggested by Oddershede {\it et
  al.}~\cite{oddershede93}.

\begin{figure}
\begin{center}
\includegraphics[scale=0.33,angle=0]{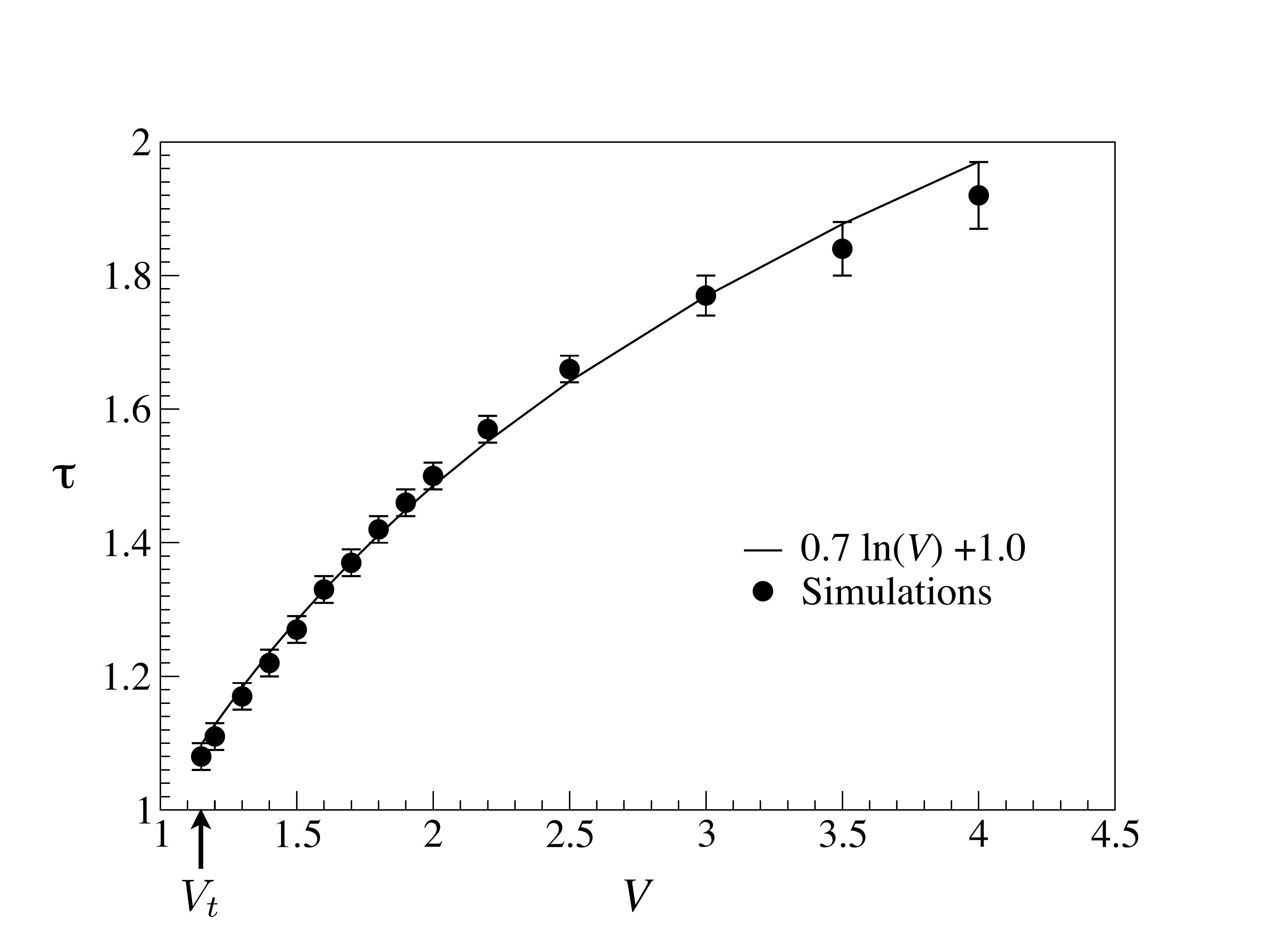}
\caption{Exponent $\tau$ of the fragment size distribution as a
  function of the impact velocity for a disc of $N=1457$
  particles. The full line corresponds to a logarithmic fit (see
  eq.~(\ref{eq.4})).}
\label{fig.4}
\end{center}
\end{figure}

As mentioned above, the fragment size distribution exhibits a power
law behaviour not only at $V_t$, but for all the values of the control
parameter greater than $V_t$. What is more, we find the exponent
$\tau$ to grow logarithmically with the impact velocity (see
fig.~\ref{fig.4}) as
\begin{equation} \label{eq.4} 
\tau = \alpha \ln{V} + \beta,
\end{equation}
with $\alpha=0.70 \pm 0.01$ and $\beta=1.00 \pm 0.01$. Because of the
decrease of $S_{\mathrm{max1}}$ with increasing $V$, the power law
region narrows, which enlarges the errors bars. Nevertheless, the
logarithmic fit is quite sensible for $V_t \le V \le 3$.

The growth of the exponent $\tau$ with the energy imparted to the
fragmenting system is clearly observed in various materials of
different shapes, like glass and plaster plates~\cite{kadono97},
ceramic tubes~\cite{katsuragi05}, gypsum disks~\cite{astrom04a} and
glass rods~\cite{ching99,ishii92}. However, as far as we know, only a
few experimental papers are concerned with the exact dependence of
$\tau$ on the energy. Thus, by performing impact experiments on rocks,
Matsui and his collaborators~\cite{matsui82} found the logarithmic
relation given by eq.~(\ref{eq.4}) with $\alpha \simeq 0.65$,
regardless of the shape and rock type investigated (granite, basalt,
tuff). Furthermore, Moukarzel {\it et al.} have very recently studied
the fragmentation of liquid (glycerol and water) droplets by a
pressurized-gas blow~\cite{moukarzel07}. As can be seen in fig. 6 of
ref.~\cite{moukarzel07}, $\tau$ has approximately a logarithmic
behaviour as a function of the jet pressure, which is related to the
imparted energy. The increase of $\tau$ with energy reported in other
simulations of impact~\cite{myagkov05} and explosive~\cite{ching99}
fragmentation, seems to be in agreement with eq.~(\ref{eq.4}), at
least at low impact energy.

In contrast, by performing the same kinds of numerical experiments,
other authors reported a virtually constant value of $\tau$ with
increasing impact energy~\cite{kun99,diehl00,araripe05,behera05},
corresponding to a zero value of $\alpha$. Also, as mentioned above,
simple analytical~\cite{meibom96,kadono02,astrom04a} and numerical
models~\cite{inaoka97,astrom04b} predict a value of $\tau$ which
depends only on the dimension of the space.

Considering these experimental and theoretical results, we suggest
that the logarithmic dependence of the exponent $\tau$ on the impact
energy (or velocity) given by eq.~(\ref{eq.4}) is generic of
fragmentation processes. A question which naturally arises is on what
parameters the coefficient $\alpha$ depends.

\begin{figure}
\begin{center}
\includegraphics[scale=0.33,angle=0]{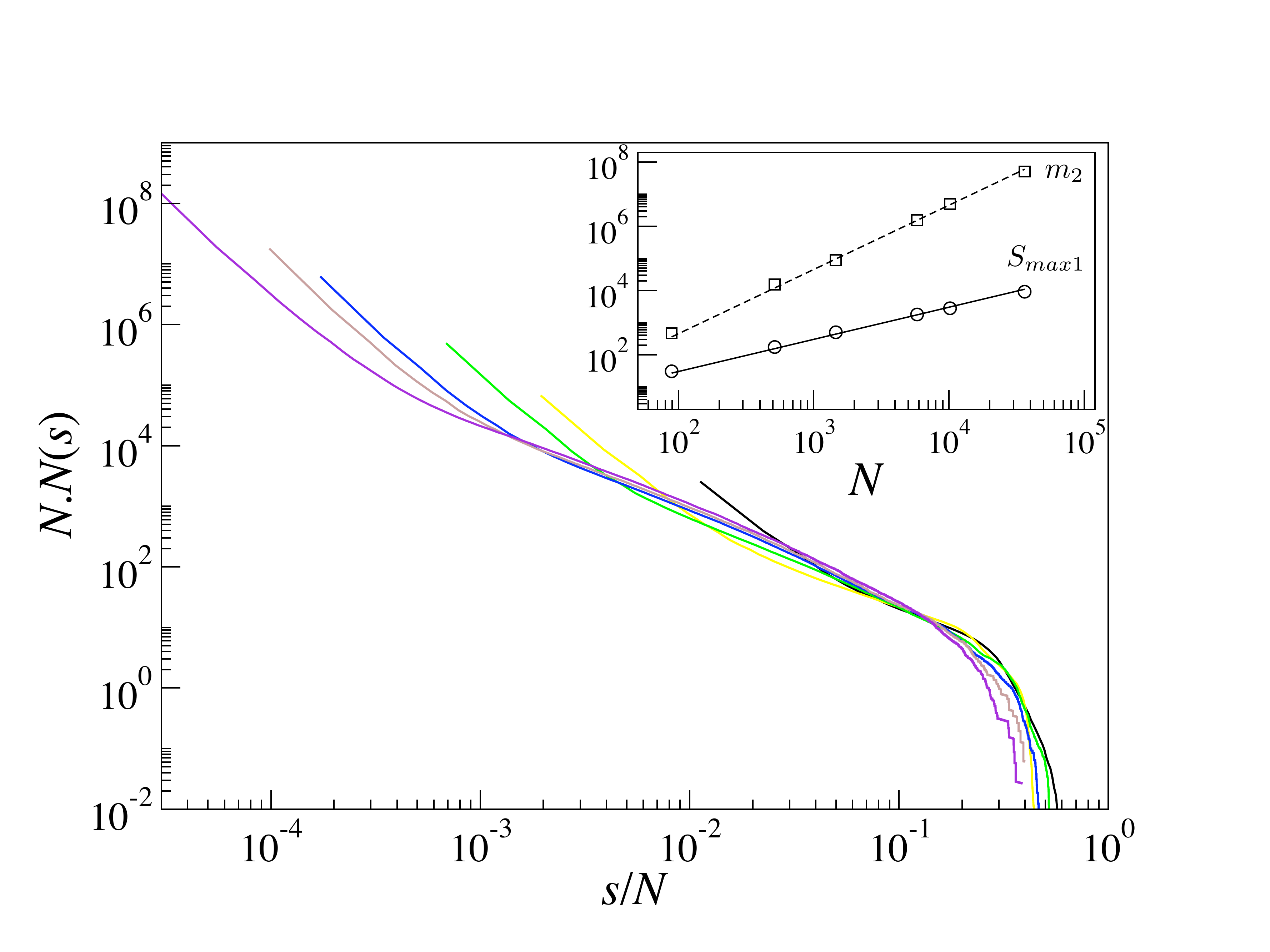}
\caption{(colour online) Scaling collapse of the fragment size
  distribution at $V=2$. From right to left $N=89$, 515, $1\;457$,
  $5\;813$, $10\;181$, $36\:289$. In the inset: $S_{\mathrm{max1}}$
  and $m_2$ are fitted by $N$ (solid lines) and $N^2$ (broken
  lines) respectively.}
\label{fig.5}
\end{center}
\end{figure}

We now turn to the finite-size scaling properties of the fragment size
distribution. It follows from the conservation of the total number of
particles that:
\begin{equation} \label{eq.5}
N  = \sum_{s=1}^{N} s \;n(s).
\end{equation}
Suppose that $n(s)\sim f(N) \; s^{-\tau}$, where $f$ is a function of
$N$ to be determined. By taking the continuous limit (large $N$), we
have
\begin{equation} \label{eq.6}
N  \sim  f(N) \int_{1}^{N} s^{1-\tau}\; ds  \sim f(N) N^{2-\tau},
\end{equation}
as long as $\tau <2$. Hence, $f(N)\sim N^{\tau-1}$, and the fragment
size distribution scales as:
\begin{equation} \label{eq.7} 
n(s) \sim N(s) \sim \frac{1}{N} \bigg( \frac{s}{N}\bigg)^{-\tau}.
\end{equation}
The fragment size distribution $N(s)$ defined by eq.~(\ref{eq.2}) is
plotted in fig.~\ref{fig.5} for various sizes $N$ of the disc as a
function of $s/N$. Considering the important surface effects at small
$N$, the data collapse is good, except for the small fragment sizes,
which do not follow a power law with exponent $\tau$. We find that the
quality of the data collapse is better at higher impact velocities.

The validity of eq.~(\ref{eq.7}) is confirmed by the scaling behaviour
of the second moment of the fragment size distribution given by
eq.~(\ref{eq.3}). Using eq.~(\ref{eq.7}), we obtain (with $\tau<3$):
\begin{equation} \label{eq.8}
m_2 \sim  N^{\tau-1} \int_{1}^{N} s^{2-\tau}\; ds \sim N^{2}.
\end{equation}
As illustrated in the inset of fig.~\ref{fig.5}, the scaling behaviour
predicted by eq.~(\ref{eq.8}) is in very good agreement with the
simulation data. Furthermore, the size of the largest fragment scales
nicely with the total number of particles, that is $S_{\mathrm{max1}}
\sim N$. It must be emphasized that the scaling behaviour of
$S_{\mathrm{max1}}$ and $m_2$ is independent of $\tau$, and
is then observed for all the velocities larger than $V_t$ investigated
in this work.

\section{Discussion}

To summarize, this simple model provides a general framework for
investigating the various parameters that may play a role in
fragmentation processes. As we have seen, our results do not depend on
the range of attraction of the interaction potential, which suggests a
certain universality in fragmentation of solid bodies. Qualitatively,
the propagation of cracks and the crack pattern obtained in our
simulations are akin to the ones observed experimentally in platelike
objects.

The power law behaviour of the fragment size distribution and the
maximum of its second moment seem to provide evidence that
fragmentation is a continuous phase transition, similar to a
percolation transition, as suggested by Kun and
Herrmann~\cite{kun99}. However, in contrast to continuous phase
transitions, the power law is observed, not only at the threshold
velocity, but also at all velocities above $V_t$. Moreover, the
scaling behaviour of the fragment size distribution is simply inferred
from the conservation of the number of particles and does not reveal a
critical behaviour.  We note that these properties are very similar to
the avalanche phenomena associated with hysteresis loops, in which
Sethna and his collaborators~\cite{perkovic99} found scaling and power
law behaviours in a large region near the critical point. However,
this analogy has to be confirmed by thorough analysis.

Furthermore, we find that the exponent of the fragment size
distribution increases as a logarithmic function of the imparted
energy, in agreement with experiments on rocks and on liquid
droplets. This result casts some doubts upon the existence of
universality classes in fragmentation, in the strict sense of
continuous phase transitions, but it sheds light on the scattering of
the values of $\tau$ measured experimentally. Besides, we expect this
logarithmic relation to be generic in fragmentation phenomena.

\vspace{2cm}

We thank H.J. Herrmann for helpful conversations, P. Viot and
H. Hietala for critically reading the manuscript.

\end{document}